# Neural Stochastic Block Model
# & Scalable Community-Based Graph Learning

Zheng Chen
College of Computing & Informatics, Drexel University

Xinli Yu
Department of Mathematics, Temple University

Yuan Ling
Amazon Alexa AI

Xiaohua Hu
College of Computing & Informatics, Drexel University

## ABSTRACT

This paper proposes a novel scalable community-based neural framework for graph learning. The framework learns the graph topology through the task of community detection and link prediction by optimizing with our proposed joint SBM loss function, which results from a non-trivial adaptation of the likelihood function of the classic Stochastic Block Model (SBM). Compared with SBM, our framework is flexible, naturally allows soft labels and digestion of complex node attributes. The main goal is efficient valuation of complex graph data, therefore our design carefully aims at accommodating large data, and ensures there is a single forward pass for efficient evaluation.

For large graph, it remains an open problem of how to efficiently leverage its underlying structure for various graph learning tasks. Previously it can be heavy work. With our community-based framework, this becomes less difficult and allows the task models to basically plug-in-and-play and perform joint training. We currently look into two particular applications, the graph alignment and the anomalous correlation detection, and discuss how to make use of our framework to tackle both problems. Extensive experiments are conducted to demonstrate the effectiveness of our approach.

We also contributed tweaks of classic techniques which we find helpful for performance and scalability. For example, 1) the GAT+, an improved design of GAT (Graph Attention Network), the scaled-cosine similarity, and a unified implementation of the convolution/attention based and the random-walk based neural graph models.

## Categories and Subject Descriptors

H.2.8 [**Database Management**]: Database Applications – Data Mining

## 1. Introduction

Graph community structure is an important research area for graph learning that receives a long-standing attention from a wide spectrum of scientific studies [1]. A graph community can be understood as a group of strongly interrelated nodes that distinguish themselves from the rest of the graph. The task of *community detection* is to discover such groups from the graph. With community detection, each graph node will be assigned a community label, also referred to as the node's *membership*. If a node can belong to multiple communities, then the node has soft community label, i.e. a $K$-dimensional probability distribution where $K$ is the number of communties in the graph, and we can say it is *mixed membership* [2, 3].

In terms of graph topology, nodes in the same community tend to interconnect more often than nodes outside of the community [4]. In terms of node attributes, nodes from the same community are usually more similar [5, 6]. The existence of rich community structures is especially recognized for large real-world networks [7, 8]. However, to our best knowledge, few previous researches have investigated how to leverage them to benefit other graph learning tasks. The entity labels for many real-world relational data, are generally not available or expensive to acquire; we believe the underlying structure of a graph is the valuable resource we can tap into.

The community structure can be beneficial from several perspectives. It can be viewed as a 'divide-and-conquer' strategy. **For performance**, the hypothesis space for a large graph learning problem could be gigantic, and we should guide the algorithm to search in the places more likely to have the right answers. For example, in the graph alignment problem in Section 3.5, nodes from two very different communities can still have similar attributes, especially when rich data like texts are not available; thus we can guide the alignment model to only match nodes from top-aligned communities to reduce the chance of mismatch. **For efficiency**, no matter it is social network, a knowledge graph or a biological network, the nodes outside a community intuitively tend to make less impact when a model learns a community. Hence, guiding the model to focus on the community and its most related communities and ignore others can help reduce unnecessary computation. **For learning**, community detection can improve the embedding quality without further demand on the data. Link prediction has been involved in graph embedding training in many researches [9-11]; if we view link prediction as the "micro" learning that helps encode graph topology, then the community detection can be viewed as the "macro" learning that equips the embeddings with better awareness of the whole picture. An example is shown in Figure 1, where it is embeddings of a graph of ten noisy communities, and the embeddings trained with community detection are clearly more separable.

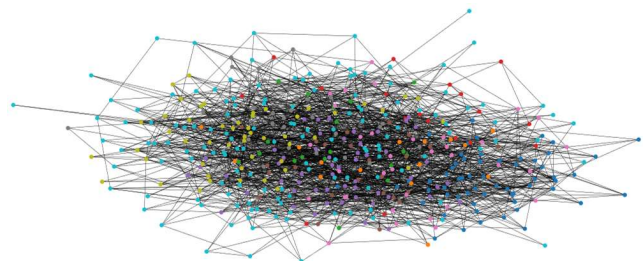

(a). a subgraph of the Amazon dataset of 10 noisy communities.

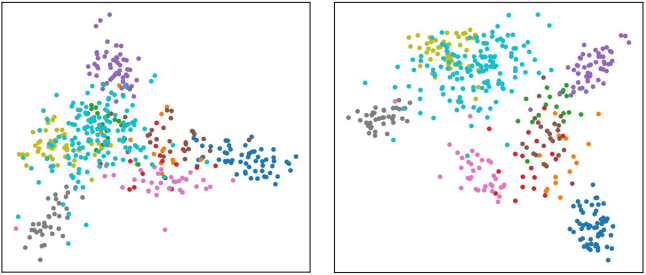

(b). embeddings trained by only the link prediction loss.

(c). embeddings trained with community detection by our joint SBM loss in (11).

**Figure 1** Illustration of the effect of community detection. Node color in (a) corresponds to embeddings of the same color in (b) and (c). Both trained with our framework in Figure 3, where the only difference is whether community detection is consider in the loss. More separation among communities is clearly seen in (c).

In this paper, we propose to train a graph learning task with our community-based framework in Figure 3. The framework learns the graph topology through the joint task of community detection and link prediction by optimizing the proposed Neural Stochastic Block Model (NSBM) layer and the joint SBM loss function in equation (11). The NSBM layer and its loss are adapted from the likelihood function of the Stochastic Block Model (SBM) [12]. In comparison to the classic SBM likelihood, the adapted loss is differentiable, naturally allows mixed-membership and weighted graph, and easily encodes node attributes. The community labels and embeddings are then passed to the task layer for task-specific use. Meanwhile, our design aims at the following requirements for online learning and scalability:
1) community label evaluation should take a single GPU-enabled feed-forward pass within the framework without the need for further iterations like the evaluation of SBM.
2) capable of accepting batches throughout the framework during the training, so that the model can learn at scale;

**Application 1**. After introduction the framework, we apply it to leverage community structure for the large graph alignment problem in Section 3.5. *Graph alignment* can be roughly described as matching corresponding nodes in different graphs so that a given *alignment cost function* is minimized [13]. We choose this task because graph alignment and community detection share similar principals at the high level: both closely rely on node similarities and topology similarities. We propose to take the convenience of the community embeddings learned by our framework and first align the detected communities before actually matching the nodes. We show in experiments that a simple neural alignment task layer on top of our framework can achieve better performance in comparison to a recent alignment model [14-16].

**Application 2**. We also apply the framework to the *anomalous correlation detection* problem in Section 3.6, which raises alert when a non-trivial subset of the data points have unusually high correlation, and we also want to pin-point this unusually correlated data subset [17-20]. The traditional way of detection is by *Principal Component Analysis* (PCA), finding the top (normalized) eigenvalue of the correlation matrix, raising alert if the eigenvalue exceeds a threshold, and finding anomalies by their similarities with the principal component. For large dataset, this approach is not just expensive, but also ineffective because the eigenvalue will be provably converge to the neutral value 0.5 as the data size grows [17]. We treat the data points as nodes of a weighted graph with their Pearson's correlations as the edge weights, and apply our framework to directly identify "communities" in that graph, and then further check each of them if they have strong internal correlations with much less computation cost.

The **main contribution** of this paper is that we develop a novel and general community-based neural graph learning framework that can learn from large-scale data and make efficient evaluation, and we show through extensive experiments that it helps improve performance for two applications. For this paper, the online evaluation time is a focus; in an industrial pipeline, the evaluation has to be efficient, while training can be off-line and take more time. We also contributed a few techniques we found helpful for performance and scalability; for example, a unified implementation of convolution-attention based and the random-walk based neural graph models, the GAT+, the *scaled-cosine similarity* to prevent embeddings from being far apart in the Euclidean space, a design to allow joint community detection and task training, etc.

## 2. Related Works

*Stochastic Block Model* (SBM) is a well-studied probabilistic generative model for graph community detection with strong theoretical background [12, 21, 22]. A further improvement of this model introduces mixed membership by adding a Dirichlet prior to the community labels [2]. The other popular framework with strong theory is the spectral clustering [23]. In comparison to SBM, the difficulty of the later is the costly computation of the eigenvectors and the requirement of availability of the entire adjacency matrix, and it becomes tricky for large graphs [24].

The probabilistic models are good for theoretical soundness and interpretability; however, they suffer from some crucial problems: 1) it is tricky to develop a complex probabilistic model than a neural network; for example, merely enabling mixed membership with a Dirichlet prior is enough to render the model intractable [2], and it has to either use the slow Gibbs Sampling, or an approximation like mean-field approximation that introduces counterintuitive independence assumptions; 2) slower inference, since iterations are typically still needed during evaluation; work-arounds like KNN is unable to work for complex data; 3) there lacks a mature general software for swift experiments and fast computation; it is often the case that a small gap in the probabilistic assumptions leads to big difference in the inference implementation.

Network alignment can be applied to identify similar users in different social networks [25] [26], study protein-protein interaction [27-29] or chemical compound network [30], align knowledge bases [31], and some computer vision tasks [32], among others. There is recent interest for large graph alignment. FINAL-N in [15] makes use of both topology and node attributes, however, it takes quadratic time and space complexity and can run out of memory on a 50K graph with 10 attributes. The other method UniAlign in [16] completes in one iteration, and supports high-dimensional features, however, the topological adjacency information is lost in the conversion of the original network to a node-feature bipartite network which results in low accuracy, especially on a large network. The algorithm still needs quadratic time complexity although it is non-iterative [33]. A more recent work [14] applies an approximation technique by landmark node embeddings, where see the concept of "landmark embeddings" are analogous to the community embeddings but obtained by pre-defined rules. However, the choice of landmarks are tricky and the performance is susceptible to the choice; the paper's proposal of "random sample" does not work well for large graphs; we also find it has to dramatically increase the number of landmarks in order to maintain reasonable performance.

Anomaly detection is also an active research topic of long history. Many algorithms focus on detecting individual anomalies [34], some other researches aim at discovering *anomaly groups* [35-39]. The individuals in each anomaly group need not appear to be anomalous by themselves, but collectively they become different from others. This paper focuses on discovering a group of strongly correlated data points with *the assumption of weak background correlation*, i.e. the majority of the normal data points should not correlate. The primary technique for CAD is PCA-based [19, 40-43]. Although CAD involves streaming data since the beginning, none of the above-mentioned works can directly apply to large-scale data due to the principal score degeneration [17]. Meanwhile, as far as we know, the best time complexity reported from previous works is sub-cubic [19, 44, 45], still intensive for large-scale monitor and involving much unnecessary calculation that can be avoided.

## 3. Model & Inference

Main notation in this paper are summarized in Table 1. By convention, small bold letters represent vectors and big bold letters represent matrices.

| General | |
|---|---|
| $G = (V, E)$ | The graph, its node set and edge set. |
| **A** | The graph adjacency matrix. |
| **X** | The row-major node embedding matrix. |
| **X̄** | The row major embeddings of a community. |
| $\mathcal{L}(\mathbf{x}) = \mathbf{Wx} + \mathbf{b}$ | A neural linear layer, and mathematically an affine transformation, where **W** and **b** are trainable parameters; $\mathcal{L}(\mathbf{X})$ is applying the linear layer to every column of **X**. |
| SBM Neural Framework | |
| $K$ | The number of communities to detect. |
| **P** | The stochastic block matrix in the SBM model. |
| **z** | The membership vector in classic SBM. |
| **Z** | The relaxed mix-membership matrix in our neural framework. |
| **C** | Records the number of edges inside a community or between every two communities for classic SBM. Relaxed as community similarity in the neural framework. |

**Table 1**. Notations

### 3.1. Stochastic Block Model

We fist give a brief overview of the canonical Stochastic Block Model (SBM), which aims to divide the nodes of a graph $G = (V, E)$ into $K$ clusters (pre-defined, and each cluster is also referred to as a block) by modeling generation of the graph edges $E$. Basically it uses links for weak supervision, constructing a likelihood for the observed links, and iteratively switches node membership to increase the likelihood, until node label convergence. Let $|V|$ be the size of the node set, and $|E|$ be the number of edges, then the model has the following parameters,

1) A membership vector **z** of size $|V|$ is labeling the membership of each vertex. For convenience, given a node $v \in V$, we can view **z** as a function and let the label of $v$ being denoted as $\mathbf{z}(v)$.
2) A $K \times K$ matrix named stochastic block matrix **P** s.t. $\mathbf{P}(i,j)$ is the probability of a vertex in community $i$ having an edge with a vertex in community $j$. If $i = j$, then $\mathbf{P}(i,j)$ is the probaibility for the intra-community edges; if $i \neq j$, then $\mathbf{P}(i,j)$ is the probability inter-community edges.

Let $C_k, k = 1, \ldots, K$ be the set of all nodes in community $i$ and let $n_k = |C_k|$. The number of all possible edges from community $i$ to community $j$ is $n_{i,j} = \begin{cases} n_i n_j & i \neq j \\ \binom{n_i}{2} & i = 1 \end{cases}$ for undirected graph and $n_{i,j} = \begin{cases} n_i n_j & i \neq j \\ 2\binom{n_i}{2} & i = 1 \end{cases}$ for directed graph. Let $e_{i,j}$ be the number of actual edges between community $i$ and community $j$. Note all $n_{i,j}$ and $c_{i,j}$ are functionss of **z**. We can in addition let $\mathbf{C} = (c_{i,j}), \mathbf{N} = (n_{i,j})$ be two $k \times k$ matrices to store above-mentioned values, and let $\mathbf{n} = (n_i)$ be a vector of length $K$. SBM infers the above parameters by maximizing the following likelihood $\ell$ of the observed edges $E$,

$$\ell = \mathbb{P}(E|\mathbf{z}, \mathbf{P}) = \prod_{\substack{v_1, v_2 \in V \\ i = \mathbf{z}(v_1), j = \mathbf{z}(v_2)}} \mathbf{P}(i,j)^{c_{i,j}} (1 - \mathbf{P}(i,j))^{n_{i,j} - c_{i,j}} \quad (1)$$

In simple words, the likelihood (1) represents a generative process for edges: for every pair of nodes $v_1, v_2$, if $v_1$ is from the $i$th community, and $v_2$ is from the $j$th community, then there is a probability $\mathbf{P}(i,j)$ that $v_1, v_2$ have an edge between them. The maximum likelihood for $\mathbf{P}(i,j)$ can then be solved as $\mathbf{P}(i,j) = \frac{c_{i,j}}{n_{i,j}}$, and the optimization for **z** goes as the following:

---
**Algorithm 1:** Canonical SBM optimization
**Result:** node community labels **z**
while **z** not converged **do**
    for each $v \in V$ **do**
        choose $i \in \{1, \ldots, k\}$ s.t. $\mathbf{z}(v) = i$ maximizes the likelihood
        update **C**, **N** and $\mathbf{P}(i,j) = \frac{c_{i,j}}{n_{i,j}}$
    end
end

---

We have several problems for equation (1) and Algorithm 1. First, (1) is non-differentiable and optimizing it involves integer programming; thus, it is not straightforward to integrate (1) with a neural network. Second, they also do not consider node attributes; although there are recent development like *mixed-membership SBM* [46, 47], the probabilistic inference inevitably grows complicated with added parameters. Third, like many other probabilistic models, a trained SBM may still need inference that takes iterations to adequately evaluate new nodes in the graph given the limited information encoded by **P**.

### 3.2. Neural SBM Community Detection

We still consider a graph $G = (V, E)$, and now also with the node attributes. For convenience, we assume the attributes of a node can be numerically represented by a $d$ dimensional feature vector. Let these feature vectors be collectively denoted by a $|V| \times d$ matrix $\mathbf{X}^1$. We start by fixing a fundamental requirement in our design. As mentioned in the introduction, we want a single feed-forward GPU-enabled pass to infer community labels during evaluation. To achieve this, the categorical labels **z** in the canonical SBM is relaxed as soft assignments (a.k.a. soft labels, or mixed membership) **Z**, a $|V| \times K$ matrix, where $\mathbf{Z}(v, i)$ is the probaiblity of node $v$ being a member of the $i$th community. Soft assignment is a natural in neural networks with the softmax layer [48]; also it avoid the expensive integer programming. Then we want to design a neural network $f$ such that,

$$\mathbf{Z} = f(V|E, \mathbf{X}) \quad (2)$$

That is, the neural network $f$ is a function that assigns a mixed membership to each node given the node features **X** and the edges $E$. For example, let **A** be the adjacency matrix (or any other appropriate matrix representing $E$, e.g. graph Laplacian), and **b** be a learnable bias, a concrete $f$ can be as simple as (3).

$$\mathbf{Z} = \text{softmax}(\mathbf{AX} + \mathbf{b}) \quad (3)$$

As another essential component, we need a loss function. If we happen to have some ground-truth community labels, then a loss like the negative log-likelihood can be constructed based on **Z** like any neural network. We denote such loss as the $\text{loss}_{\text{labels}}$.

---

3    [1] This follows recent neural network researches and tools like Tensorflow or PyTorch that treat a data point as a row vector. In this case a linear layer (an affine function) is written as $\mathbf{xW}^T + \mathbf{b}$ rather than $\mathbf{Wx} + \mathbf{b}$. Conventionally computational science and machie learning instead treat a data point as a column vector.

We should address the case when we only have edges for weak supervision without ground-truth community labels. We turn the likelihood in equation (1) as the loss function. By first applying a logarithm on $\ell$, and then applying the following approximation (4) for the logarithmic terms,

$$\ln(1-x) = -x\ln x + x + r, \forall |x| < 1 \quad (4)$$

where $r$ is a infinitesimal remainder such that $\lim_{x\to 0} r = 0$; taking logarithm of (1) and then replace the logarithmic terms by (4), with several careful mathematical transformations, we will arrive at an approximate likelihood written in the matrix format as in (5) where $\circ$ denotes element-wise product, and **1** is a vector with all components being 1. See the foot-note sketch proof..

$$\ell \approx \mathbf{1}^T(\mathbf{C}\circ \ln \mathbf{C})\mathbf{1} - (\ln \mathbf{1}^T\mathbf{Z})\mathbf{C}\mathbf{1} - \mathbf{1}^T\mathbf{C}(\ln \mathbf{Z}^T\mathbf{1}) \quad (5)$$

We note that (5) holds for the majority of real-life networks, because real-life networks and their communities tend to obey the densification laws [49]. We define $\text{loss}_{\text{SBM}} = -\ell$ as the *SBM loss*. The SMB loss favors stronger links inside a community, but weaker links among communities, and hence encourages diagonally dense $P$ as showing in Figure 2. A rigorous proof is not hard and hence omitted due to limited space.

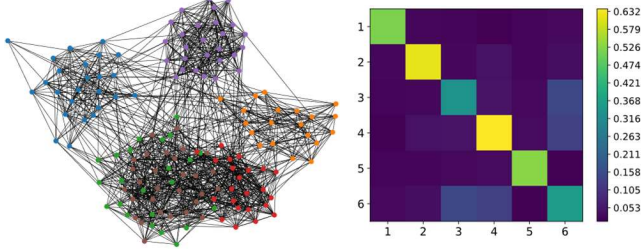

(a) the case when every node is one of the $K = 6$ communities; the right plot is the colormap of its $6 \times 6$ stochastic block matrix **P**.

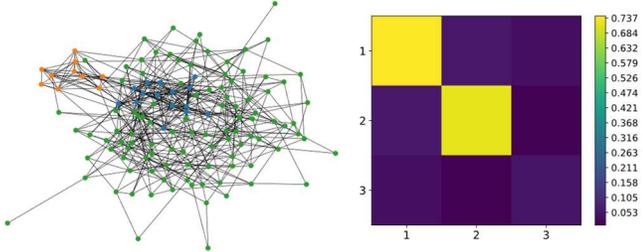

(b) the case when only some nodes belong to the $K = 2$ communities; the right plot is the colormap of its $3 \times 3$ stochastic block matrix **P** where a "pseudo-community" is added to hold other nodes not in the two communities.

**Figure 2** Illustration of two community detection cases of our concern.

It is possible not every node belongs to a community with stronger internal connections. The application in Section 2.6 is such a case, where there are one or more anomaly clusters, but remaining non-anomalies lacks strong links. Our solution is to add a *pseudo-community* to hold nodes not clustered into the $K$ communities, as illustrated in Figure 2 (b). Let **Z** be a $|V| \times (K+1)$ matrix so that every node has an additional option that it is not in any community; the stochastic block matrix **P** also gets one additional dimension. However, we should not encourage nodes to fall in this pseudo-community, thus the pseudo-community is not considered in **C** and $\text{loss}_{\text{SBM}}$.

We are not there yet. Although **Z** is now relaxed to allow decimals, **C** is not, so (5) is still not differentiable. By the definition in Section 3.1, **C** holds the counts of edges between every two communities in its off-diagonal cells, and the number of intra-community edges along its diagonal. We notice the fact that the counts in $\mathbf{C}(i,j)$ is larger if community $i,j$ have more connections, which bears strong resemblance to "community similarity". Therefore, we generalize the matrix **C** as the *community similarity matrix.* From this generalized perspective, a **C** can be

$$\mathbf{C} = \mathbf{Z}^T(\mathbf{XX}^T + \mathbf{AA}^T)\mathbf{Z} \quad (6)$$

Moreover, as we can see from equation (6), thinking of **C** as similarities naturally allows the edges being weighted, or even associated with edge features. Based on our experiments, we find it essential to consider two additional loss terms. Forthgoing, we denote a linear layer (mathematically an affine function) as $\mathcal{L}(\mathbf{x}) = \mathbf{xW}^T + \mathbf{b}$ where **W** is the trainable feature weights and **b** is the trainable bias.

1) the link prediction loss - given two nodes $v_1, v_2$ and denote their embeddings by $\mathbf{x}_1 = \mathbf{X}(v_1), \mathbf{x}_2 = \mathbf{X}(v_2)$, the link prediction can be made by $\sigma\big(s(\mathcal{L}_1(\mathbf{x}_1), \mathcal{L}_2(\mathbf{x}_2))\big)$, where $\mathcal{L}_1, \mathcal{L}_2$ are linear layers, and $\sigma$ is an acativation function like the sigmoid function, $s$ is a similarity metric like dot product, cosine similarity, a differentiable norm, or any other appropriate function. $\mathcal{L}_1, \mathcal{L}_2$ can be the same if the graph is undirected. Then a general form of the link prediction loss is written as

$$\text{loss}_{\text{link}} = \sum_{\substack{v_1, v_2 \in V \\ x_1 = \mathbf{X}(v_1), x_2 = \mathbf{X}(v_2)}} l\Big(\sigma\big(s(\mathcal{L}_1(\mathbf{x}_1), \mathcal{L}_2(\mathbf{x}_2))\big)\Big) \quad (7)$$

where $l$ is a function turning a prediction score into a loss, e.g. negative log-likelihood in the case of unweighted graph.

$$l(\sigma) = \begin{cases} -\log(\sigma) & (v_1, v_2) \in E \\ -\log(1-\sigma) & (v_1, v_2) \notin E \end{cases} \quad (8)$$

For the metric $s$, in particualr, we find a simple but effective similarity metric, which is rarely seen in other researches to our best knowledge. We find it work better than all other similarity functions mentioned above in our tweaking experiments,

$$s(\mathbf{x}_1, \mathbf{x}_2) = \alpha \times \cos(\mathbf{x}_1, \mathbf{x}_2) \quad (9)$$

Here $\alpha$ is a sufficiently large constant, and by parameter tuning we find $\alpha = 16$ generally works well. We call (9) as the *scaled cosine similarity*. We can view it as being enforcing similarity normalization to prevent large separation in Euclidean space if using dot product, but allows sufficient gradients to flow back to optimize the parameters.

2) let $\mathbf{Z}(v)$ be the mixed membership of a node $v$, then the entropy loss (regularization) like (10) will encourage the major probability mass being assigned to only one of the labels, and penalties a diverse distribution.

$$\text{loss}_{\text{entropy}} = \sum_{v \in V} \text{entropy}\big(\mathbf{Z}(v)\big) \quad (10)$$

Although being a regularization, we find the entropy loss essential to our purpose. The SBM loss optimizes for the macro sparsity among different communities, while the entropy loss optimizes for the micro sparsity between nodes and communities; together they ensures discovery of reasonable community boundaries.

Overall, we sum all four losses discussed above, where "$\text{loss}_{\text{labels}}$" is optional. We call this the *joint SBM loss*. We call the neural module defined in this section as the *Neural Stochastic Block Model* (NSBM).

$$\text{loss} = \text{loss}_{\text{SBM}} + \text{loss}_{\text{entropy}} + \text{loss}_{\text{link}} + \text{loss}_{\text{labels}} \quad (11)$$

### 3.3. Combining with Neural Graph Embedders

The previous section has setup a relatively complete SBM neural network. This network can be placed on top of a neural

---

**Sketch proof of (5)**.

Taking logarithm of (1) then it becomes $\ell = \sum_{i,j \in \{1,\dots,K\}} \big(c_{i,j}\ln c_{i,j} - n_{i,j}\ln n_{i,j} + (n_{i,j} - c_{i,j})\ln(n_{i,j} - c_{i,j})\big)$; then use (4), the likelihood will be transformed to $\ell = \sum_{i,j \in \{1,\dots,K\}} c_{i,j} \ln \frac{c_{i,j}}{n_{i,j}} - |E| + O\left(\frac{c_{i,j}}{n_{i,j}}\right)$; then note $n_{i,j} = |C_i||C_j|$, where $|C_i|$ is the number of nodes in the $i$th community, and plugin this, to $\ell$, we will have $\ell \propto \sum_{i,j \in \{1,\dots,K\}} c_{i,j} \ln c_{i,j} - \sum_{i=1}^{K} c_{i,*} \ln|C_i| - \sum_{j=1}^{K} c_{*,j} \ln|C_j|$ where $c_{i,*} = \sum_{j=1}^{K} c_{i,j}$ and $c_{*,j} = \sum_{i=1}^{K} c_{i,j}$; this is the same as (5) when written in terms of matrices. We can prove $O\left(\frac{c_{i,j}}{n_{i,j}}\right) \to 0$ for a graph with constant densification power.

graph embedder which generates the embeddings **X** to plugin equation (3). Although there is also possibility to plugin edge embeddings into (6) as mentioned earlier, for this paper we focus on the node embeddings. The architecture is shown in Figure 3.

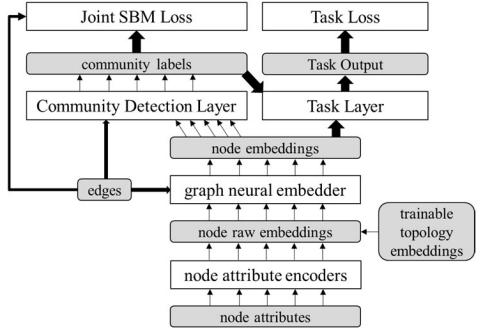

**Figure 3** The proposed community-based graph learning framework.

1) The node attributes are encoded by attribute encoders. For example, the text attributes can be encoded by the pretrained-BERT [50], the categorial or numerical data can be encoded by a feed-forward network. The outputs of different encoders for one node as well as an additional trainable embedding vector are concatenated as a single raw embedding vector.
2) Then we input the raw embeddings into a neural graph embedder, for example, the Graph Convolutional Network [51], the Graph Attention Network [52], or node2vec [9]. The output are node embeddings.
3) On one hand, the node embeddings are sent to a community detection layer (e.g. equation (3)) to generate community labels, which are then optimized through the joint SBM loss as defined in (11). On the other hand, the node embeddings can be passed to a task module to accomplish some specific tasks, whose results are optimized through the task loss. In this paper, we later study two tasks, the network alignment problem, and the anomalous correlation detection problem.
4) In addition, the results of community detection can feed the task module, either as additional features, or to potentially divide the hypothesis space of the task for the purpose of "divide-and-conquer" as discussed in introduction.

Let **X** also denote the set of all node embeddings in the graph, and let $\widetilde{\mathbf{X}}_1, \dots, \widetilde{\mathbf{X}}_K$ denote *community node embeddings*. Since the community label is soft, then intuitively $\widetilde{\mathbf{X}}_k = \text{diag}(\mathbf{Z}(:,k)) \mathbf{X}$. In practice, we clip **Z** by a threshold $\theta^{(\mathbf{Z})}$ (e.g. 0.1). We can optionally re-normalize the clipped **Z**, but in practice we find it not effective.

Often there is need in the task module for a *community embedding* $\widetilde{\mathbf{x}}_1, \dots, \widetilde{\mathbf{x}}_K$ for each community, like our applications in later Section 3.5 and Section 3.6. Let $\tilde{f}$ denote the function that assembles node embeddings in a community as a single community embedding,

$$\widetilde{\mathbf{x}}_i = \tilde{f}(\widetilde{\mathbf{X}}_i), i = 1, \dots, K \quad (12)$$

For example, if the output embeddings of the neural graph embedder have encoded necessary local graph topology and attributes, then letting $\tilde{f}$ be a weighted sum given all embeddings in $\widetilde{\mathbf{X}}_i$ then we find it generally sufficient to apply the following self-attention weights, where $\mathcal{L}$ is a linear layer,

$$\text{self\_atten}(\widetilde{\mathbf{X}}_i) = \text{softmax}\left(\widetilde{\mathbf{w}}_2^\text{T} \tanh\left(\mathcal{L}(\widetilde{\mathbf{X}}_i)\right)\right)$$
$$\widetilde{\mathbf{x}}_i = \widetilde{\mathbf{X}}_i \left(\text{atten}(\widetilde{\mathbf{X}}_i)\right)^\text{T}, i = 1, \dots, K \quad (13)$$

This design will be directly used for Section 3.5 the graph alignment task. For Section 3.6 anomalous correlation detection, a variant (19) without the softmax is used.

**How this design works for joint training**. Modern tools like Tensorflow or PyTorch support the clipping function. If the community node embeddings or the community embeddings are used in the task module, information will flow back from the task loss to the community detection parameters, and hence we achieved joint training of the community detection and the task.

### 3.4. Adaptation for Large Data

The design of equation (3) and (6) as well as some neural graph embedders requires availability of adjacency matrix and all node embeddings, preventing our framework in Figure 3 to train on a large graph. The current publicly available implementations of neural graph models like the graph convolutional networks (GCN) [51], or the graph attention networks (GATs) [52], they both use sparse matrices to store the adjacency-related matrices. However, at this moment, sparse matrix operations are still not well-supported by major deep-learning tools like Tensorflow or PyTorch; the sparse matrix will be blown into a full matrix when multiplied by a dense matrix, a very possible situation during backpropagation, preventing its application to a complex model and large data. In this section, we discuss our approach to adapt our framework to large data.

First, we propose that, many current popular neural graph embedders can be "almost" equivalently implemented in the same way as sequence embedding, and hence no need for sparse matrices. This is straightforwardly true for random-walk based methods like deep-walk or node2vec. For GCN, its convolutional layer derived by Chebyshev polynomial approximation can be viewed as applying pre-defined and fixed "attention" weights on neighborhood nodes; and two stacked convolutional layers can be achieved by running two epochs with one convolutional layer; for GATs, it is the same but now the attention weights are learnable self-attentions. Therefore, we propose that such neural graph embedders in practice can be implemented using the same architecture as illustrated in Figure 4.

1) It requires pre-computing node representatives $\text{repr}(v)$ for each $v \in V$, typically including $v$ itself, plus its neighborhood nodes, or nodes along a sample path. The nodes in $\text{repr}(v)$ are sorted by a criterion. We use the following sorting key for nodes from neighborhood, which always puts $v$ as the first node in the sequence.

$$|\deg(u) - \deg(v)|, u \in \text{repr}(v) \quad (14)$$

For nodes from a sample path, we could just use the node order in the sample path.

2) The raw embeddings of the nodes in $\text{repr}(v)$ are passed through a sequecne embedder, like the Transformer, which further transforms the raw node embeddings, as well as encoding the sequence order. The output embeddings of the sequence embedder are then combined as a single node embedding vector by a sequence-to-vector layer, e.g. average pooling, max pooling or self-attention. Optionally, we may train the sequence embedder with a negative-sampling based language-modeling task, e.g. the skip-gram in node2vec.

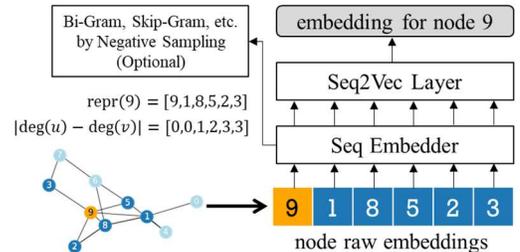

**Figure 4** The general architecture for seq-based neural graph embedder, consisting of a sequence embedder, a sequence-to-vector layer, and optionally a language-modeling module by negative sampling. Here we choose (14) as the representative node sorting key; other sorting keys like Jaccard index are also applicable.

Then, since the mixture of node embeddings have happened in the neural graph embedder, it becomes unnecessary to place adjacency information in (3), therefore we reformulate it as (15), where $\mathcal{L}$ is a linear layer.

$$\mathbf{Z} = \text{softmax}(\mathcal{L}(\mathbf{X})) \tag{15}$$

Last, we deal with (6), by proposing the following scheme to construct training batches compatible with the sequence-based neural network training in Figure 4.

1) We maintain a list recording each node's *major community assignment*, i.e. the community of the highest membership probability. We initialize this list by a heuristic algorithm, starting with a node with high degree, and then keeps "absorbing" closely linked adjacent nodes into the same community, until no such node can be found from the neighbors. The detailed algorithm is omitted for space. Then we update the list at the end of every batch based on the output of (14).

2) Then construct training batches according to Algorithm 3. We first randomly choose from the communities, and then random sample a batch without replacement from the chosen communities. For each batch, the computation of (6) and the joint SBM loss in (11) are now constrained to that batch.

---
**Algorithm 2:** Initial community assignment by degree-based expansion.

**Result:** training batches
**Parameters**:
*batch_size*, the size of each training batch; for example, 256
$c$, the number of communities to sample each time
*major_comm_assign*: a list of major community assignment for nodes
**Process**:
1. Update corresponding major community assignment at the end of the forward pass of each batch.
2. Randomly sample $c$ communities.
3. Randomly sample nodes of *batch_size* without replacement from the nodes of the sampled communities.

---

### 3.5. Graph Alignment

For two graphs $G_1 = (V_1, E_1), G_2 = (V_2, E_2)$, our alignment task is to match every node in $G_1$ with one node in $G_2$. If we allow different graph sizes, and without loss of generality let $|V_1| \leq |V_2|$, then the alignment can be formulated as the following optimization,

$$\min_{\mathbf{P}} \left( d(\mathbf{A}_1, \mathbf{P}\mathbf{A}_2\mathbf{P}^{\mathrm{T}}) \right) \tag{16}$$

where $d$ is some distance measure, like the Frobenius norm, $\mathbf{A}_1, \mathbf{A}_2$ are the adjacency matrices of the two graphs, and $\mathbf{P}$ is a $|V_1| \times |V_2|$ *selection matrix* (every row has one element being 1 and all others being 0, no two 1s are in the same row or same column). The meaning of (16) is treating each row of the adjacency matrix as the feature of a node, and $\mathbf{P}$ is exactly an operator that chooses one node from $G_2$ for each node in $G_1$; thus $\mathbf{P}$ represents the alignment, and is our optimization target. The $\mathbf{P}$ can be relaxed to allow every row being a discrete distribution to avoid integer progamming, similar to what we did for the community labels in Section 2.2. To handle node embeddings in (16), we have to place corresponding embedding vectors in non-zero cells of $\mathbf{A}_1, \mathbf{A}_2$ (so they become 3-D tensors).

The above formulation cannot handle large graphs due to involvement of adjacency information of the whole graph for globally optimized alignment. We consider that, if the graph topology could be well captured in the node embeddings due to other modules of our framework, i.e. the neural graph embedder, the SBM loss, the link-prediction loss, etc., then it could be less necessary for a global optimization. Therefore, we reduce the graph alignment task to a much simpler task - embedding similarity search. For fair comparison experiment, we perform embedding similarity search by the same k-d tree algorithm given the node embeddings $\mathbf{X}_1, \mathbf{X}_2$ of $G_1$ and $G_2$ [14]. Using a more mature package like Faiss [53] is another option.

However, there is a hidden problem - $\mathbf{X}_1, \mathbf{X}_2$ may not be comparable. The two graphs are embedded separately, and no parameters in our framework offer to encode their comparability. We thus introduce the alignment loss with entropy regularization, where $\mathcal{L}_1$ and $\mathcal{L}_2$ are two linear layers.

$$\text{loss}_{\text{align}} = d(\mathcal{L}_1(\mathbf{X}_1), \mathbf{P}\mathcal{L}_2(\mathbf{X}_2)) + \text{entropy}(\mathbf{P}) \tag{17}$$

The alignment $\mathbf{P}$ can now be further modeled as the result of a neural alignment layer by (18), which shares the same linear layers in (17), and again the $s$ is a similarity measure like in (7),

$$\mathbf{P} = \text{softmax}\left(s(\mathcal{L}_1(\mathbf{X}_1), \mathcal{L}_2(\mathbf{X}_2))\right) \tag{18}$$

We note it is essential to have the linear layers $\mathcal{L}_1$ and $\mathcal{L}_2$; their job is exactly to transform embeddings in $\mathbf{X}_1$ and $\mathbf{X}_2$ to a comparable space.

Now the training objective is reduced to learn an *alignment projection* $\mathbf{P}$ that maps embeddings from two graphs to a comparable space by optimizing (17), it is not necessary to train on all embeddings together, i.e. we can train by batches. We still follow Algorithm 3 to construct batches, but with the following extra operations.

1) **Compute community embeddings** by (13).
2) **Compute community similarity** between $G_1$ and $G_2$ by applying the loss (17) on community embeddings.
3) **Compute node embedding alignment**; for $c$ communities sampled from $G_1$, sample the same batch size from their top-$c$ matched communities in $G_2$, and then applying the loss (17) again on node embeddings.

There is a caveat that we also train the alignment projection $\mathbf{P}$ on community embeddings; we construct batches of community embeddings and put them at the end of each epoch, so that $\mathbf{P}$ also learn to project community embeddings in order to perform above step 2). The community detection module is inactive for these special batches. After all training, we pass embeddings of $\mathcal{L}_1(\mathbf{X}_1)$ and $\mathbf{P}\mathcal{L}_2(\mathbf{X}_2)$ to the embedding similarity search algorithm.

### 3.6. Correlated Anomaly Detection

Given a $n \times d$ feature matrix $\mathbf{X}$, where $n$ is the number of data points and $d$ is the dimension of data features, then the *correlated anomaly detection* is to find $K$ subsets of of $\mathbf{X}$, such that each subset has a non-trivial number of features, and the correlation within each such subset is high. Each such subset is called a *correlated anomaly set*. Until now, this task is mostly solved with variants of *Principal Component Analysis* (PCA) across the $n$ features. Let's also denote the $K$ feature subsets as $\widetilde{\mathbf{X}}_k$ for convenience, then this task has two parts,

1) Find the feature subsets $\widetilde{\mathbf{X}}_k, K = 1, \ldots, K$ with strong internal correlations.
2) Raise alarm when the *principal score* of a feature subset $\widetilde{\mathbf{X}}$ exceeds a certain threshold. The score is defined as $\rho(\widetilde{\mathbf{X}}) = \frac{\lambda_{\max}(\text{corr}(\widetilde{\mathbf{X}}))}{\text{size}(\widetilde{\mathbf{X}})}$ (normalized top eigenvalue of the data correlation matrix) and $\rho(\widetilde{\mathbf{X}}) \in (0, 1]$. A high principal score is an indicator there is strong correlation within $\widetilde{\mathbf{X}}$.

The approach taken here is non-conventional. We aim to solve this problem through our NSBM framework in Figure 3. The idea is to treat the correlation matrix as an adjacency matrix of a weighted graph. Note again the design in (6) naturally allows weighted adjacency matrix.

**For community detection**, we let the number of communities be the same as the number of anomaly sets, i.e. $K$, so that each community corresponds to one anomaly set. We add one extra pseudo-community for holding non-anomalies like in Figure 2 (b). The "edge" between two features $\mathbf{x}_1, \mathbf{x}_2$ is represented by the clipped correlation, where $\theta^{(\text{corr})}$ is the clipping threshold. We set an emperically rule to find this threshold, e.g. 1.5 times of the average correlation of historical data.

$$\text{corr}'(\mathbf{x}_1, \mathbf{x}_2) = \begin{cases} 0 & \text{corr}(\mathbf{x}_1, \mathbf{x}_2) < \theta^{(\text{corr})} \\ \text{corr}(\mathbf{x}_1, \mathbf{x}_2) & \text{otherwise} \end{cases}$$

**For the task**, we use (13) without softmax as unnormalized attention weights; here $\widetilde{\mathbf{X}}_k$ is $\mathbf{Z}$-augmentd as in (13)

$$\boldsymbol{\alpha} = \widetilde{\mathbf{w}}_2^T \tanh(\widetilde{\mathbf{W}}_1 \widetilde{\mathbf{X}}_k + \tilde{\mathbf{b}}) \quad (19)$$

and then the following is the community embedding for the $k$th community,

$$\tilde{\mathbf{x}}_k = \widetilde{\mathbf{X}}_k \boldsymbol{\alpha} \quad (20)$$

Let $\text{var}(\cdot)$ denote the variance, and we then apply the following specialized loss, which we refer to as the *PCA loss*, on top of each community embedding $\tilde{\mathbf{x}}_k, k = 1, \ldots, K$, based on a theorem that PAC is equivalent to such optimization [54],

$$\text{loss}_{\text{PCA}}^{(k)} = \text{var}(\tilde{\mathbf{x}}_k) + 50(\boldsymbol{\alpha}^T\boldsymbol{\alpha} - 1)^2 \quad (21)$$

By minimizing this loss, the following should approximate the principal score for the $i$th community,

$$\rho(\widetilde{\mathbf{X}}_k) \approx \frac{\boldsymbol{\alpha}^T(\text{corr}'(\widetilde{\mathbf{X}}_k))\boldsymbol{\alpha}}{\text{size}(\widetilde{\mathbf{X}}_k)} \quad (22)$$

**For evaluation**, given a feature matrix, compute its clipped correlation matrix and treat it as if it was a weighted graph; the features go through the graph embedder and get turned into "node" embeddings, and then the community detection module labels each feature with categorical labels $1, \ldots, K + 1$, where $1, \ldots, K$ means a feature likely belongs to a correlated anomaly set, and label $K + 1$ means the feature is not an anomaly. For each of the $K$ anomaly set, compute its approximate principal score $\rho$ by (22) and raise an alarm if $\rho$ is higher than an alarm threshold $\theta^{(\text{anomaly})}$ (e.g. 0.7).

## 4. Experiments

### 4.1. Community Detection

We first show our framework's capability of discovering community structures before moving into specific applications. We experiment on the following datasets with ground-truth labels. Note for all these datasets, a node may belong to multiple ground-truth communities (i.e. the communities can overlap).

|  | Youtube | Amzn Prod | Wiki PageLinks |
|---|---|---|---|
| # nodes | 21477 | 5072 | 756K |
| # comms | 237 | 301 | 451 |
| comm-size | 50~2217 (med. 90) | 20~328 (med. 51) | 1K~418K (med. 1.5K) |

**Table 2** A summary of community detection data sets. "med.": median.

1) The Youtube dataset contains a social network of users, where each community is a ground-truth group on Youtube [55]. Only node adjacency is available for this dataset. We consider
2) The Amazon Co-Purchase dataset [56] provides product categories as the ground-truth labels. It is a graph such that, if a product is "also-bought" with another product at least three times, then there is an undirected edge between them. Each product is associated with the product title.
3) The Wikipedia page-link dataset is a web graph of Wikipedia hyperlinks [57]. In additional to the graph topology, every node is associated with the title of the wiki page. The page category is the ground-truth community label.

We propose assemble a Transformer on top of the graph attention network GAT as designed in Figure 4, and we call this as *GAT+*. We equip our framework in Figure 3 with GAT+ and node2vec, referred to as the NSBM-GAT+ and NSBM-n2v. We compare with the GAT+ and n2v without the SBM loss to show the effectiveness that loss function. A 100-dimensional vector is used to encode the topology information for a node. For the Youtube dataset, it is just this topology embedding, and no attribute. For the Amazon and Wikipedia dataset we compare the case with, the attribute encoder is summarized in equation (23); we apply the pre-trained BERT (base) plus a linear layer to turn a title into a 100-d text embedding, which is concatenated with the topology embedding to form a 200-dimensional node embedding, and then they go through one additional linear layer.

$$\text{attr\_emb}(title) = \mathcal{L}_2([\mathcal{L}_1(\text{BERT}(title)), \mathbf{x}]) \quad (23)$$

For unsupervised learning, we compare with n2v and GAT+ with the link-prediction loss (but still feed the same output from the attribute encoders). We take the unweighted average of precision and F1 score (a.k.a. marco-F1) across the communities. In case when one node belongs to multiple (say k) communities in the ground-truth, the prediction is good for the community if that community is ranked top-k in the model's soft prediction label. The experiment results are shown in Table 3. This shows the effectiveness of the joint SBM loss by a large margin. Also we notice adding text data dramatically improves the especially the precision and hence the F1 score.

We also test our framework's ability when trained with partial labels, shown in Figure 5; we in addition compare with a neural classification model LINE. We allow the models to see a random sample of ground-truth labels during training. The results indicate the joint SBM loss provides helpful information especially when the labels and text are less seen. When more labels or the texts become available, they start to dominate the training, and the metric numbers converge.

|  | Youtube | | Amzn Prod (without/with text) | | Wikilinks (without/with text) | |
|---|---|---|---|---|---|---|
|  | a. pre | ma. F1 | a. pre | ma. F1 | a. pre | ma. F1 |
| n2v | 0.308 | 0.457 | 0.35/0.54 | 0.46/0.56 | 0.34/0.47 | 0.24/0.49 |
| GAT+ | 0.332 | 0.469 | 0.37/0.56 | 0.47/0.58 | 0.37/0.48 | 0.25/0.50 |
| NSBM-n2v | 0.398 | 0.530 | 0.43/0.61 | 0.53/0.62 | 0.39/0.52 | 0.29/0.54 |
| NSBM-GAT+ | **0.412** | **0.531** | **0.45/0.65** | **0.54/0.64** | **0.41/0.54** | **0.29/0.56** |

**Table 3** Compare the joint SBM loss in (11) with the link-prediction loss when trained without labels. Due to varying community sizes, we take unweighted average of metrics across the communities; "a. pre" is the average precision, and "ma. F1" is the macro-F1. For both n2v and GAT+, the SBM loss optimize the model for a clear better understanding of graph topology in terms of community detection.

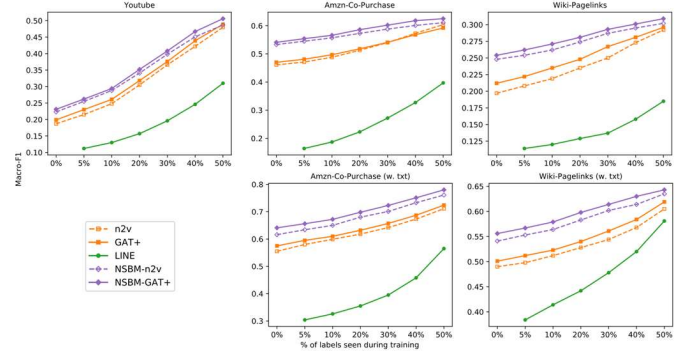

**Figure 5** Performance comparison between our framework (SBM-n2v, SBM-GAT+) and GAT+, n2v and LINE when some labels are available. The advantage of applying the joint SBM loss is obvious when less labels are seen for all datasets tested.

## 4.2. Network Alignment

We apply our NSBM framework to the network alignment problem as discussed in Section 3.5. We compare with several recent models UniAlign [16], FINAL [15], CAlign [58] and REGAL [14]. We experiment with four datasets with different characteristics summarized in Table 4 to test our framework's generalized performance. The metric is the alignment accuracy - the total number of correct alignment predictions over the size of the ground truth. For UniAlign and REGAL, we supply count vectors for text tokens (removing stop words and very rare words). The neural models can directly feed on the original text.

We first follow previous researches to test how the above methods perform with different levels of *pre-alignment*. For UniAlign and REGAL, we merge the features of two pre-aligned nodes. FINAL has a build-in support for pre-alignment. For CAlign, we fix pre-alignment through the iterations. For NSBM, we let the alignment module see the alignment labels and train by negative-sampling classification. Results are shown in the top row of Figure 6.

We then check if each method is capable of effective and efficient alignment for new nodes coming into the graph. We hide 50% of the nodes with ground-truth alignment during training, and divide them into 5 batches for evaluation, and measure the average performance; this is simulating online and zero-shot evaluation. The measurement of evaluation time includes node attribute encoding. For UniAlign, we treat the other 50% aligned during the training as pre-alignment, and then make the evaluation. For node2vec, we sample from adjacent nodes' pre-sampled path nodes, which saves time and prevents the time measure being dominated by path sampling. For NSBM, we limit PyTorch to use only one CPU for fairness. Results are shown in the bottom row of Figure 6.

| name | Aminer-LinkedIn | *Amzn Co-Purc* | *Fliker-Last.fm* | *DBLP* |
|---|---|---|---|---|
| # nodes | 30K/34K | 37K/40K | 22K/26K | 390K/400K |
| overlap | 4K (13%) | 30K (78%) | 641 (3%) | 180K (46%) |
| text | yes | yes | no | no |
| type | mixed | product graph | social network | academic |

**Table 4** A summary of four alignment data sets.

1) *ArnetMiner-LinkedIn*. The ArnetMiner coauthor dataset comes with co-author relations and affiliation, job, research keywords. The LinkedIn network comes with the "also viewed" data to create a "co-viewed" network, and every person has skills, education background and career history. The ArnetMiner network contains 30,045 nodes; the LinkedIn network contains 34,221 nodes. There are 4,269 nodes from the two graph that can be identified as the same person due to name match and large keyword overlap. We break affiliation/location information into categorical node attributes, and use the research key words and the LinkedIn interests as the text data. We see from the results that our approach is superior to other methods. UniAlign and FINAL do not properly handle this dataset. Due to the sparsity of the alignable nodes in the dataset, REGAL based methods are impaired by the randomly sampled landmark nodes, which do not necessarily provide related information for the alignable nodes. **For NSBM**, it directly and fully encodes node attributes and graph topology for superior performance. The zero-shot evaluation has similar performance results. NSBM's evaluation time cost is reasonable, two to three times higher than REGAL.

2) *Amazon Co-Purchase Network* [56]. The first graph comes from year 2015 and contains 37,201 nodes; the second network comes from year 2016 and contains 40,021 nodes. The two networks have 29,760 nodes in common based on product ID. We use product title words as the text data. This dataset differs from the ArnetMiner-LinkedIn dataset on that it is two homogenous networks, and they have large overlap. As a result, performance for most methods are much better. **For NSBM**, the model generally achieves best results and is able to generalize good performance to zero-shot evaluation, but the accuracy does considerably drop since a substantial amount of nodes are hidden during training and the node embedding quality is hence reduced.

3) *Fliker-Last.fm*. These are two social networks, and the links in the graph represent friendship. The first network has 21,945 nodes; the second network has 26,426 nodes. It has a ground truth of 641 matched nodes. Only user names are available. We measure the Levenshetein distance and longest common substring between each node and its neighbors, normalize both metrics by the average length of two user names, discretize both into three categories and then make counts as the categorical node features. This is again a heterogenous network alignment with sparse overlap. **For NSBM**, it achieves reasonable competitive results considering both accuracy and time.

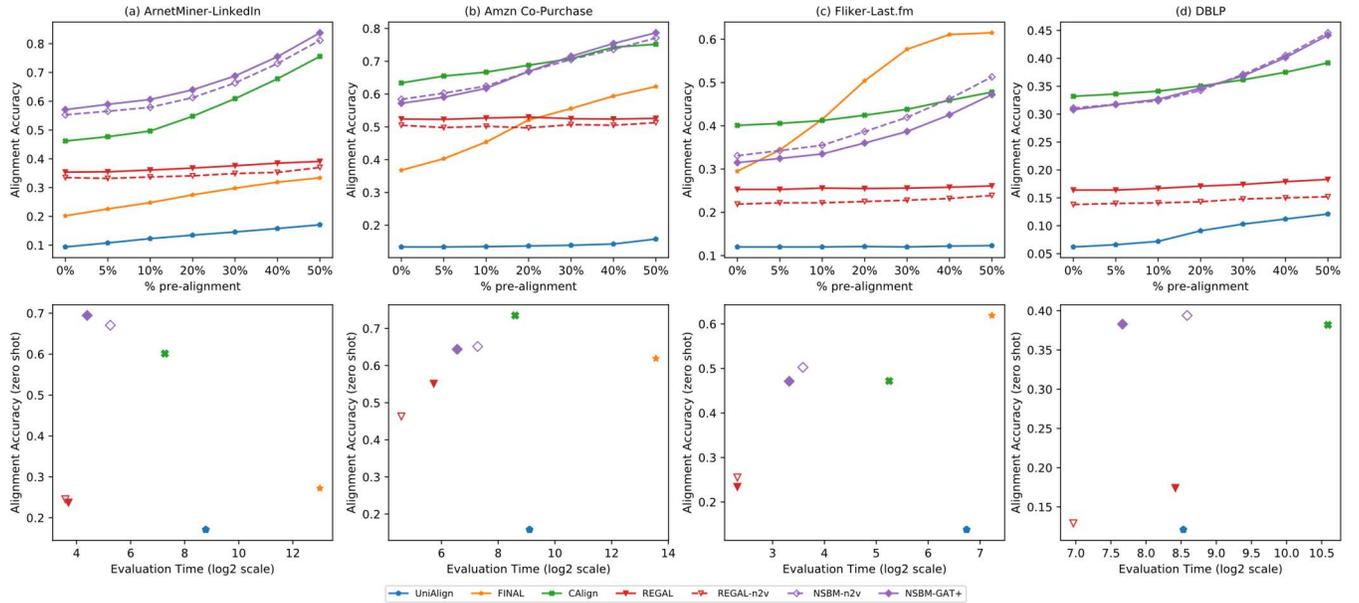

**Figure 6** Network alignment performance comparison. First row: alignment accuracy at different percentages of alignment labels being used for training. Second row: zero-shot online evaluation accuracy vs. evaluation time (seconds, log2 scale); 50% of alignable nodes in the first graph is hidden from training; our two NSBM models are at the top-left corner of the plots, indicating it is generally more efficient and more effective.

4) *DBLP*. The 2015 and 2016 DBLP co-author network [59]. Only the graph is available. The 2015 graph has 389,809 nodes, and the 2016 graph has 402,040 nodes, and they have an overlap of size 179,519. This dataset is to test the models learning capability for large graph without any extra attributes. This dataset is too large for FINAL, and for UniAlign we need fast randomized SVD, but the performance is inadequate. For CAlign, its evaluation takes long time. For REGAL, the impact of such a large graph is clear; it is underperforming in comparison to other dates sets. We also checked using ten times of the landmark nodes, which can get the performance above 0.4, but in that case the evaluation time cost is near half hour and any number of landmark nodes is not growing logarithmically with the graph size. **For NSBM**, it has a distinctive upper-hand performance for this dataset, confirming our capability for a large graph. The GAT+ and n2v versions perform almost the same in this dataset. We find GAT+ learns an attention that favors large-degree nodes on this dataset, while node2vec favors large-degree nodes by design. This makes sense since a connection to a high-degree node in the co-author network is intuitively more informative. **For** efficiency, the evaluation time significantly drops even below REGAL due to no text encoding layers.

Overall, the experiments show our NSBM framework generally achieves remarkable performance for the graph alignment task in terms of comparison with recent methods tested above. It is especially suitable when rich attributes are available. It is efficient to align new nodes while still maintaining strong accuracy. NSBM-GAT+ generally works better than NSBM-n2v when node attributes are very informative; otherwise it seems to depend on whether high-degree nodes in the graph are more important.

### 4.3. Anomalous Correlation Detection

We only consider NSBM-GAT+ in this section due to the nature of our problem – we push a sliding time window through the time series data, and monitor if the data in a window has strong unusual correlation. For NSBM, this is equivalent to test a new graph each time, and node2vec sampling is costly for this purpose. GAT+ is more efficient as it only identifies adjacent nodes based on the clipped correlation matrix. We evaluate if NSBM-GAT+ is more capable of detecting correlations in comparison to PCA. We use two data sets.

1) a large 4-month ecommerce web server log dataset of 315M Apache log entries and visits from 2M distinct IPs. The server is known to be constantly harassed by web crawlers; in this case, a node is an IP, and the edge is its correlation with other IPs in their requests for different server resources. A feature vector for this data consists of how many times an IP visits each sever resource in the given time window.
2) Daily stock price history of about 7200 U.S. stocks; in this case, a node is a stock, and the edge is a stock's price-change correlation with other stocks. A feature vector for this data consists of the daily price change (in percentage) in a given time window.

We use a sliding window on both datasets to mimic online-monitoring. There are about 20,000 windows for the server log data with window size from about 500 to 4000. There are 4,000 windows for the stock data with window size gradually growing from 400 to above 7000. In a time window, if there is a non-trivial subset of nodes with principal score (normalized top eigenvalue of correlation matrix) exceeding 0.7, we say that windows has *anomalous correlation*. Both sets satisfy the assumption of weak background correlation. Even the stocks from the same business section on average only strongly correlate in 1.5% of the time.

**Ground-truth simulation**. We need simulated anomalies due to lack of ground-truth. For the server log data, several crawlers were used to simulate realistic anomalous visits on the server. For the stock data, hundreds of good-quality sets of anomalies are manually identified, which are usually stocks from related business sections; these true anomalies are perturbated and replicated for simulation.

**Qualitative evaluation**. We first apply a synthetic-data trained NSBM to the real data and visualize the results to qualitatively compare with the PCA approach. To construct the training set, we first sample anomaly sets from the simulated data and find their mean and covariance matrix, and then generate the synthetic anomalies from normal distributes with these mean and covariance. We then make synthetic feature matrices from a normal distribution with the true-data background mean and variance, and randomly replace some of the feature vectors by the synthetic anomalies.

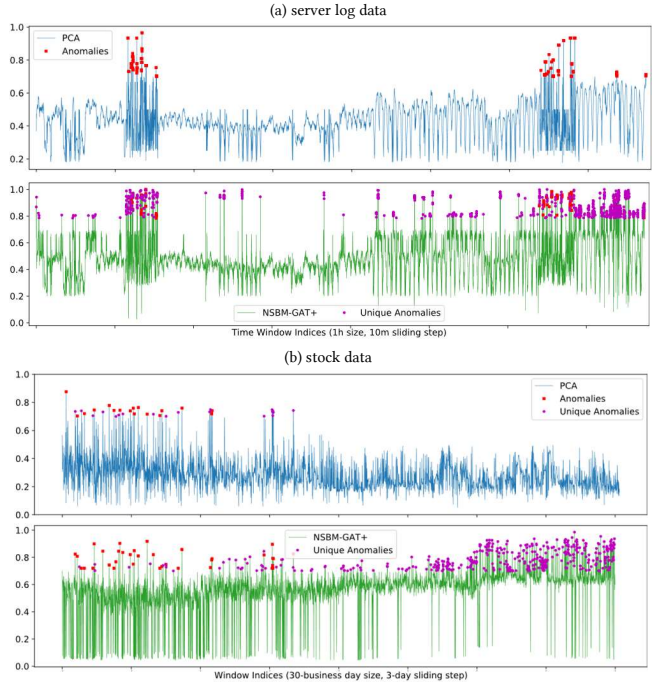

**Figure 7** Apply the synthetic-data trained NSBM-GAT+ to detect anomalous correlations from the (a) server-log data and the (b) stock data. PCA only raises alarm when the correlation is overwhelming. Especially for stock data, the number of stocks grows over time, an the PCA approach never works after a certain time. NSBM-GAT+ is capable of discovering the anomalies throughout the time.

**Quantitative evaluation**. For the following, we denote the synthetic-data trained NSBM-GAT+ as NSBM[1]. We also train with two other schemes,
1) Train with 50% of the original data, denoted by NSBM[2].
2) Randomly injects 50% of the synthetic anomalies into 50% of the time windows for semi-supervised training; the community detection module has a label loss (negative log-likelihood) that penalizes the case when these anomalies are clustered into the pseudo-community; this is denoted by NSBM[2].

To be fair, we remove anomalies identified from the real data in the previous experiment, and then randomly inject the remaining 50% of simulated anomalies not seen during training to 10% of the time windows as ground-truth. In this case, if an alarm is raised for a time window without the simulated anomalies, then it can be treated as a false alarm. We manipulate the anomalies to simulate three scenarios: 1) "large", the simulated anomalies will take 20% to 50% of data in the window; 2) "small", the simulated anomalies will take 5% to 20% of data in the window; 3) "hidden", 20~200 anomalies with low feature quantity are injected into a window of size larger than 2000. The last case is when all PCA-based approach is destined to fail. We set $K = 2$; an anomaly is detected as long as it is in the $K$ clusters. Results are shown in Table 5. For convenience, we use NSBM[1], NSBM[2], NSBM[3] to represent NSBM-GAT+ being trained in the three ways mentioned earlier.

|  | Large | | Small | | Hidden | | Overall | | |
|---|---|---|---|---|---|---|---|---|---|
|  | rec | acu | rec | acu | rec | acu | rec | acu | ex.a |
| PCA | **0.8**[*] 0.79 | **0.8**[**] | 0.46 0.38 | 0.74 | 0.03 0.04 | 0.51 | 0.45 0.65 | 0.78 | 76[***] |
| NSBM[1] | 0.75 0.50 | 0.68 | 0.52 0.42 | 0.65 | 0.53 0.44 | 0.66 | 0.60 0.47 | 0.67 | **22** |
| NSBM[2] | 0.79 0.75 | 0.80 | 0.53 0.48 | 0.81 | 0.54 0.49 | 0.72 | 0.62 0.66 | 0.79 | 40 |
| NSBM[3] | 0.80 0.78 | **0.84** | **0.54** **0.50** | **0.83** | **0.54** **0.53** | 0.72 | **0.63** **0.69** | **0.82** | 43 |
| Stock Data | | | | | | | | | |
| PCA | 0.33 0.53 | **0.85** | 0.10 0.55 | 0.73 | 0.03 0.04 | 0.53 | 0.48 0.69 | 0.82 | 20 |
| NSBM[1] | 0.69 0.47 | 0.62 | 0.66 0.45 | 0.53 | **0.65** 0.43 | 0.54 | 0.67 0.46 | 0.60 | 28 |
| NSBM[2] | 0.69 0.64 | 0.84 | 0.66 0.58 | 0.83 | 0.64 0.49 | 0.76 | **0.67** 0.62 | 0.80 | **7** |
| NSBM[3] | **0.69** **0.66** | 0.84 | 0.66 **0.61** | **0.84** | 0.63 **0.54** | 0.78 | 0.66 **0.64** | **0.84** | **7** |

**Table 5** Anomalous correlation experiment results. [*]"rec" includes the % of raised alerts of those injected with simulated anomalies, and the % of detected simulated anomalies in these alerted time windows; [**]"acu" is the % of detected anomalies are from the simulated anomalies; [***]"ex .a" is # alerted time windows without injected anomalies, an indicator of how likely the model is to raise a false alarm.

We see the PCA-based approach only works well for the "large" scenario, while the NSBM approach has strong performance for all scenarios, a result consistent with our qualitative comparison in Figure 7. Moreover, in terms of **computation time**, the average PCA runtime is 26s for the server log data and 37s for the stock data, while NSBM on average takes 5s for both datasets.

## 5. Conclusion

This paper proposes a general community-based large-graph learning neural framework with a loss function adapted from the stochastic block model, a unified graph encoder implementation, a mechanism to plugin an jointly train with a task module, and other careful designs for scalable training. We showed how the framework can be effectively applied in the graph alignment problem and the anomalous correlation detection problem, and we achieve competitive to better performance with significant save of evaluation time. The framework is a very general and we may apply it to other graph learning tasks in the future.